\documentclass[aps,physrev,preprint,groupedaddress]{revtex4-2}

\usepackage{amsmath}
\usepackage{amsthm}
\usepackage{mathtools}
\usepackage[dvipsnames]{color}					
\usepackage{amssymb}
\usepackage{hyperref}
\usepackage{multirow}
\usepackage{comment}

\newcommand{\transpose}{^\text{T}}
\newcommand{\bs}{\boldsymbol}

\begin{document}


\title{A modified particle filter that reduces weight collapse}


\author{Shay Gilpin}
\email[]{sgilpin@arizona.edu}
\affiliation{Department of Mathematics, University of Arizona, Tucson, AZ, USA}

\author{Michael Herty}
\email[]{herty@igpm.rwth-aachen.de}
\affiliation{RWTH Aachen University, Institute Geometry and Practical Mathematics, 52056 Aachen, Germany}
\affiliation{Extraordinary Professor, Department of Mathematics and Applied Mathematics, University of Pretoria, Private Bag X20, Hatfield 0028, South Africa
}


\date{\today}

\begin{abstract}

Particle filters are a widely used Monte Carlo based data assimilation technique that estimates the probability distribution of a system's state conditioned on observations through a collection of weights and particles. A known problem for particle filters is weight collapse, or degeneracy, where a single weight attains a value of one while all others are close to zero, thereby collapsing the estimated distribution. We address this issue by introducing a novel modification to the particle filter that is simple to implement and inspired by energy-based diversity measures.  Our approach adjusts particle weights to minimize a two-body energy potential, promoting balanced weight distributions and mitigating collapse. We demonstrate the performance of this modified particle filter in a series of numerical experiments with linear and nonlinear dynamical models, where we compare with the classical particle filter and ensemble Kalman filters in the nonlinear case. We find that our new approach improves weight distributions compared to the classical particle filter and thereby improve state estimates.

\end{abstract}


\maketitle

\section{Introduction}\label{sec:introduction}
Monte Carlo based data assimilation algorithms are a popular class of statistical estimation techniques that update an ensemble, or collection, of state variables with observations that are noisy and possibly sparse. One such method is the \textit{particle filter}~\citep[see e.g.,][]{gordon1993novel,doucet2001introduction,kunsch2013particle,vetra2018state,van2009particle,van2015nonlinear,van2019particle}.  
In general, particle filters update a posterior distribution conditioned on a set of observations, where this posterior distribution is estimated from a sequence of weights $\{w_i\}$ associated with an ensemble of particles $\{x_i\}$ using Bayes theorem. Many different forms have been proposed in recent years \cite[see][and references therein]{zaritskii1975monte,morzfeld2012random,vanden2013data,van2015nonlinear}.

While particle filters are simple to implement, their application is often limited to low-dimensional systems due to the phenomena known as weight collapse, or degeneracy  \citep[e.g.,][]{snyder2008obstacles,snyder2015performance,li2005curse,bickel2008sharp,morzfeld2017collapse}. Weight collapse occurs when a single weights attains a value close to one, while all other weights are nearly zero, effectively concentrating all information onto one particle and rendering the posterior distribution approximated by these weights useless. One way to avoid this collapse is to increase the number of particles, however it has been shown that the number of particles must grow exponentially with the dimension of the observations, therefore limiting its applicability in high-dimensional systems \citep{bengtsson2008curse}. Several alternative approaches have been proposed to counteract weight collapse. For example, a suitable number of particles can be chosen by estimating the effective dimension of the system, \citep{chorin2013conditions}. Other approaches include localization, which limits the impact of observation information and thereby the increase in particle weights, \citep{morzfeld2017collapse}. We refer the reader to \cite{van2019particle} for an overview of suitable alternatives.

Motivated by recent progress in the introduction of energy-based diversity measures in multi-objective optimization~\citep{coello2020evolutionary}, we propose a new modification to the particle filter that employs a similar concept to adjust the posterior weights. In this formulation, the particle filter weights are adjusted in  such a way that they correspond to a minimal configuration of a suitable two-body energy potential. We are able to obtain weight distributions by balancing their value within a potential landscape. This technique has been successfully applied in the context of particle swarm optimization methods for multi-objective minimization \citep[see e.g.,][]{borghi2023adaptive,borghi2022consensus}. This approach yields a simple modification to the classical particle filter that, and as we demonstrate in a series of numerical experiments, improves weight distributions and reduces the frequency of collapse relative to the classical particle filter. This work provides a proof of concept for this modified particle filter through a series of numerical experiments, which will be followed by analysis in future work.

This paper is organized as follows. 
We begin with the standard formulation of the particle filter in Sec. \ref{II.A}, followed by the modification of the weights in Sec \ref{II.B}. In Sec \ref{II.C} provides a motivating example to illustrate the modification's impact on weight computations relative to the classical particle filter. We conclude this section with a discussion of the ensemble Kalman filter in Sec. \ref{II.D}, which is another class of Monte Carlo estimation techniques we will compare with in our numerical experiments. 
We present two sets of numerical experiments, the first for a linear dynamical system with a known solution in Sec. \ref{III}, and the second for the nonlinear Lorenz '63 dynamical system in Sec. \ref{IV}. This is followed by a summary and discussion in Sec. \ref{V}.

\section{Monte Carlo based data assimilation: Particle filters, the modified particle filter, and ensemble Kalman filters}\label{sec:particle filters}
In this section, we introduce the standard formulation of the particle filter, which we refer to as the classical particle filter, and our modification using potential functions. This is followed by a motivating example where we illustrate that the introduced potential actually allows to shift the weight distribution of the filter. We conclude this section with a brief discussion of ensemble Kalman filters, which are an alternative class of Monte Carlo based data assimilation algorithm we will use in our nonlinear numerical experiments.

\subsection{The classical particle filter}
\label{II.A}
To introduce the standard formulation of the particle filter, we follow the presentation in \cite{van2019particle}. Consider an ensemble of $N_e$ model states $x_i \in \mathbb{R}^{N_x},$ for $i=1,\dots,N_e$ called \textit{particles}. These particles represent the empirical measure of the prior probability density function $p(x)$,
\begin{align}\label{dirac}
    p(x) = \frac{1}{N_e} \sum\limits_{j=1}^{N_e} \delta(x-x_j). 
\end{align}
Between observations, the particles are propagated (forecasted) using deterministic model $f:\mathbb{R}^{N_x} \to \mathbb{R}^{N_x}$ given by 
\begin{align}\label{eq:model dynamics}
    x^f_i = f(x_i),
\end{align}
for all $i$, which may be nonlinear. Random forcing can be added to the dynamics in \eqref{eq:model dynamics}. The observation $y \in \mathbb{R}^{N_y}$ is given by 
\begin{align}\label{eq:observation def}
    y = H(x_{true}) + \epsilon,  
\end{align}
where $x_{true}$ is the true state of the system, the operator $H\colon \mathbb{R}^{N_x}\rightarrow \mathbb{R}^{N_y}$, and $\epsilon$ is measurement, or observation, error. It is often assumed that $\epsilon$ is normally distributed with zero mean and $N_y\times N_y$ observation error covariance matrix $\mathbf{R}$, which is typically a scalar multiple of the identity matrix.

The observation $y$ and the predicted state of the system $x^f$ are assimilated using a likelihood function, i.e., the probability density $p(y|x)$ of an observation $y$ given a possible model state $x.$ The posterior probability distribution $p(x^f|y)$ is obtained using Bayes' theorem 
\begin{align}\label{bayes}
p(x^f|y) = \frac{ p(y|x^f)p(x^f) }{ p(y) }.
\end{align}
Using the representation of the prior probability for the particles $x_i^f$ for $i=1,\dots,N_e$, by \eqref{dirac}, we can approximate the posterior distribution with the ansatz as
\begin{align}\label{posterior}
    p(x^f|y) \approx \sum\limits_{j=1}^N w_i \delta(x^f_i - x^f), \quad \sum\limits_{j=1}^{N_e} w_i =1.
\end{align}
The unknown weights $w_i$ are obtained using Bayes' theorem \eqref{bayes} as 
\begin{align}\label{weights}
    w_i = \frac{ p(y|x_i^f) }{ \sum\limits_{j=1}^{N_e} p(y|x_i^f) }.
\end{align}
Note that all quantities in \eqref{weights} are known, that is, the position of the particles $x_i^f$ as well as the observation $y$, leading to an explicit computation of the weights $w_i$. This leads to an approximation of the posteriori distribution given by a weighted empirical measures \eqref{posterior}. 

The dynamical evolution of the particles in \eqref{eq:model dynamics} and the assimilation of observation information in \eqref{posterior} and \eqref{weights} define the two steps of a \textit{data assimilation cycle} with the classical particle filter \citep{van2019particle}. Repeating this cycle to assimilate more observations will result in a skewed distribution of weights $w_i$, ultimately causing one particle to have a nonzero weight while the others are zero or close to zero \citep{doucet2001introduction}. Several different algorithms have been designed to combat this degeneracy ranging in complexity, such as resampling methods, proposal densities, and others \citep[e.g.,][and reference therein]{doucet2001introduction,van2009particle,van2019particle}. In the next section, we propose an alternative method to prevent such weight collapse inspired by energy diversity measures.



\subsection{Modified particle filters}
\label{II.B}
As outlined in the introduction, we propose a simple procedure to update the weight distribution based on recent considerations for multi-objective minimization \cite{borghi2023adaptive}. The idea is to slightly modify the weights in \eqref{weights} to guarantee an equi-distribution of the weights. In order to measure the clustering of the weights, we introduce a diversity measure on the weight distribution. Consider the probability density $\varrho^{N_e} \in \mathbb{P}(\mathbb{R}) $ of the weights $\{ w_i \}$ given by the empirical measure $\varrho^{N_e}(w)=\frac{1}{N_e} \sum\limits_{j=1}^{N_e} \delta(w-w_j).$ In general, the diversity $\mathcal{U}$ of a probability measure $\varrho \in \mathcal{P}(\mathbb{R})$ is defined  by a two--body potential 
\begin{align}\label{potential} \mathcal{U}(\varrho)=\int\int U(x-y) \varrho(dy) \varrho(dx), \end{align}
where $U:\mathbb{R}\to\mathbb{R}$ is, for example, the Morse potential $U(z)=\exp(-C \| z\|)$ for some constant $C>0$. Other potentials, such as a Newtonian potential, are possible \citep[see e.g.,][]{fonseca2009evolutionary,braun2015obtaining,borghi2023adaptive}. Obtaining an equi-distribution then amounts to obtaining a configuration that is minimal with respect to the diversity measure $\mathcal{U}.$ Solving the minimization problem $\min\limits_{\varrho} \mathcal{U}(\varrho)$ on the space of empirical measures of size $N_e$ is, however, non--trivial and computationally expensive. Therefore, we follow a heuristic strategy similar to \cite{borghi2023adaptive}: We consider a parameterized family  $t\to \varrho(t)$ of probability measures $\varrho(t) \in \mathcal{P}(\mathbb{R})$, such that 
\begin{align}
    \frac{d}{dt} \mathcal{U}(\varrho(t)) \leq0, \; \varrho(0)=\varrho^{N_e},
\end{align}
and evolve these over a fixed time period. The decay of $\mathcal{U}$ can be achieved if $\varrho$ is a gradient flow with respect to $\mathcal{U}$ and fulfills weakly 
\begin{align} \partial_t \varrho + \partial_x \int U'(x-y) \varrho(t,dy) \varrho(t,dx) = 0, \; \varrho(0)=\varrho^{N_e}.\end{align}  
Note that the time scale $t$ here has no physical meaning. We, therefore, consider an explicit Euler discretization in time with a single(!) time step $\Delta t >0$. Furthermore, $\varrho^{N_e}$ is an empirical measure and therefore the evolution of $\varrho$ can be simply reformulated in an update of the initial weights $w_i.$ More precisely, we have 
\begin{align}
    w_i(\Delta t) = w_i - \frac{\Delta t}{N_e} \sum\limits_{j=1}^{N_e} U'(w_j-w_i). 
\end{align}
Denoting the arbitrary time step $\Delta t = \alpha > 0$ as our algorithmic parameter, we propose to obtain the posterior distribution \eqref{posterior} as
\begin{align} \label{posteriori modified}
 p(x^f|y) = \sum\limits_{j=1}^N w_i(\alpha) \delta(x^f_i - x^f),
\end{align}
where \begin{align} \label{modified weights} w_i(\alpha) = w_i -  \frac{\alpha}{N_e} \sum\limits_{j=1}^{N_e} U'(w_j-w_i)\end{align} 
and the weights $w_i$ are given by \eqref{weights}. If necessary, the weights $w_i(\alpha)$ are projected on $[0,1].$  By the previous considerations, our proposed method balances the weight computation according to Bayes' theorem with a gradient descent step on the diversity measure of the weight distribution. The balancing is controlled by the parameter $\alpha>0.$ In the case where $\alpha=0$, we recover \eqref{weights}.  

\begin{figure}[htb]\center
\includegraphics[width=.5\textwidth]{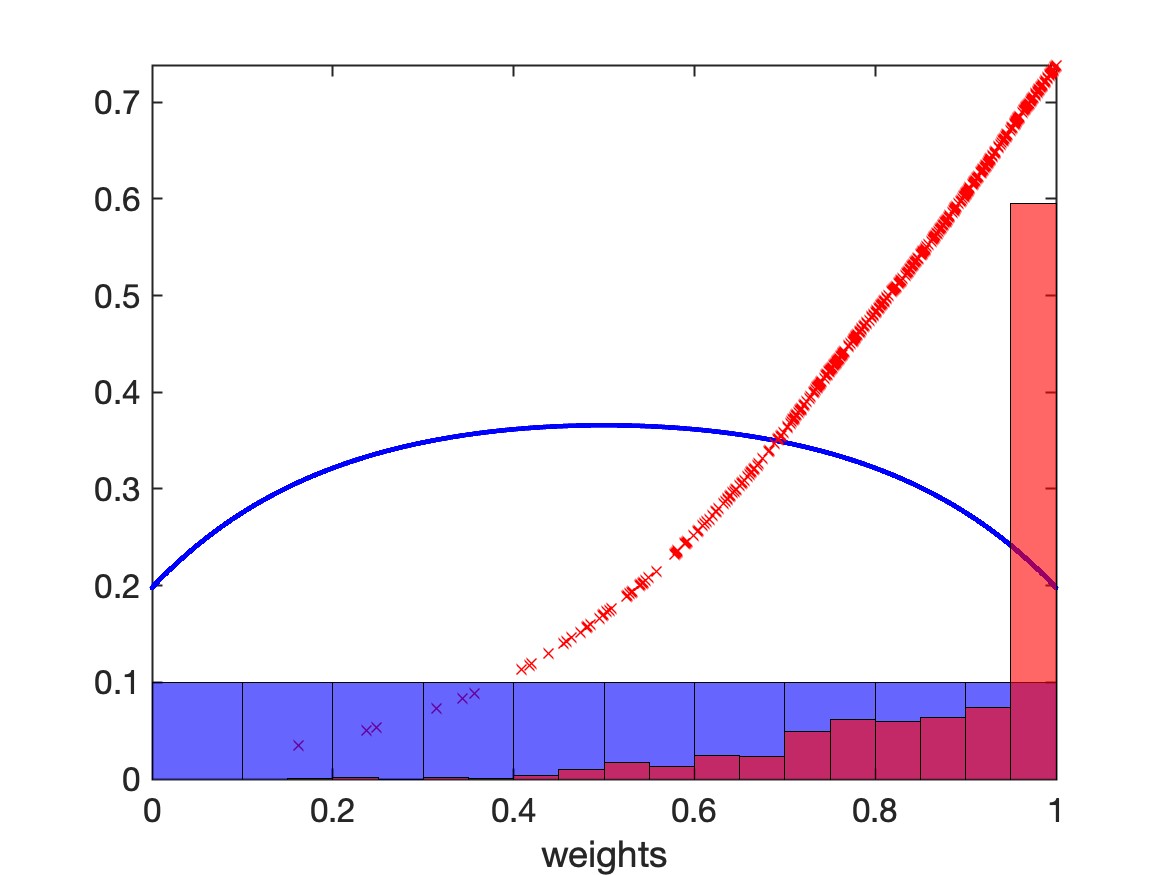}
\caption{Morse potential for an equi-distribution of weights (blue) and a distribution clustered at weight one (red). Histograms of the weight distributions are also reported. } \label{potential.fig}
\end{figure}

As an illustration, we show the Morse potential on an equi-distribution of $N_e=10^3$ weights as well as for a normal distribution of weights close to one. The histogram of the weights are given in blue and red, respectively. The corresponding values of the Morse potential $U(z)=\exp(-\frac12 z)$ are also indicated in Figure \ref{potential.fig}. It is clearly visible that the clustering leads to higher potential values, motivating a gradient descent approach for the potential as modification for the filter. 

\subsection{Motivating example}
\label{II.C}
\begin{figure}[htb]
\center
\includegraphics[width=.3\textwidth]{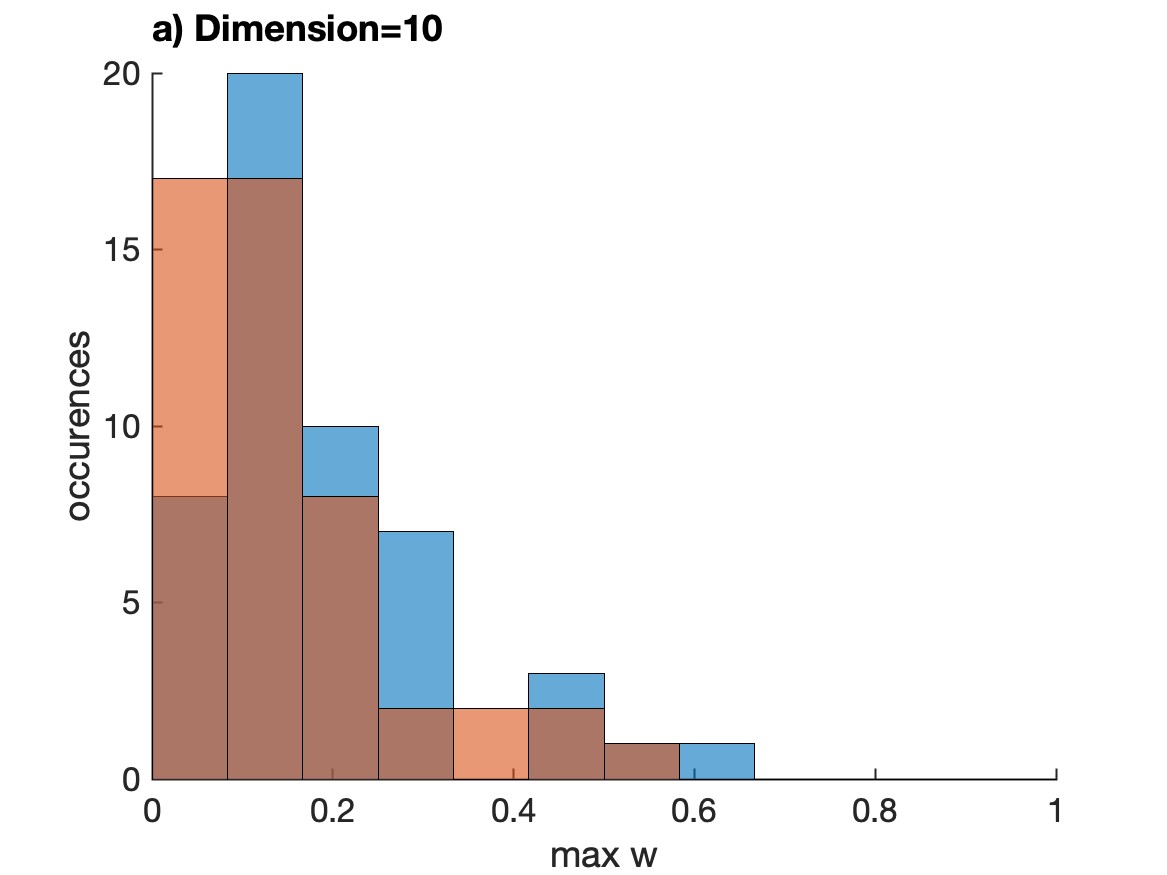}
\includegraphics[width=.3\textwidth]{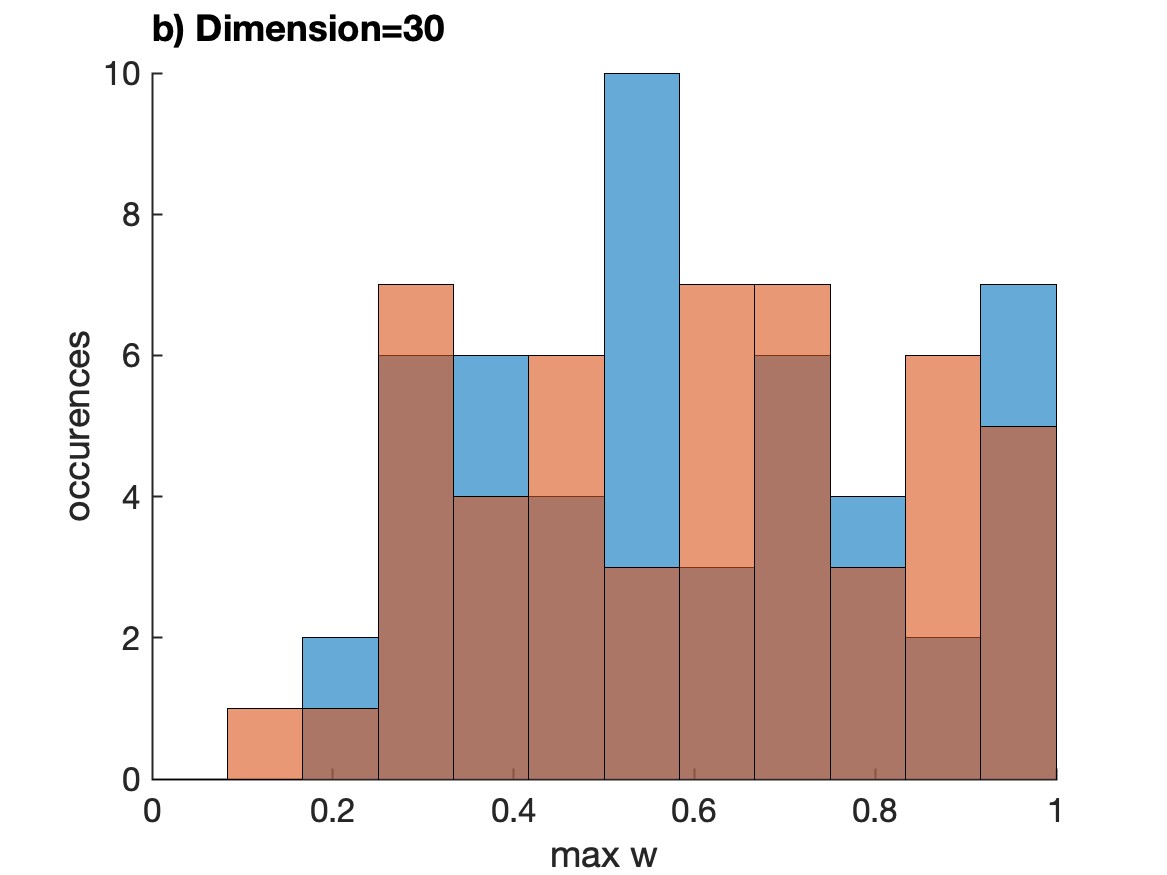}
\includegraphics[width=.3\textwidth]{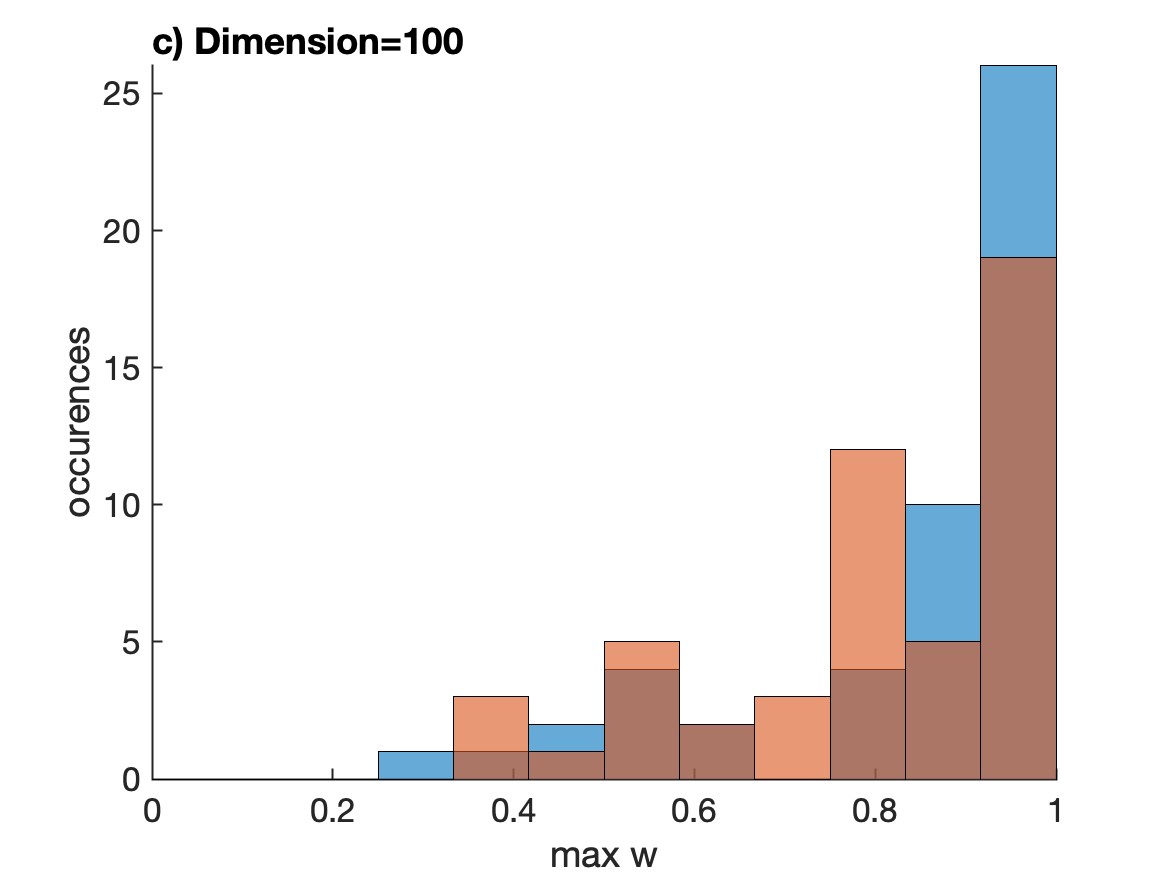}
\caption{Histogram of the maximal weights as in the example of Sec.~3 of \cite{snyder2008obstacles}. Computation of the weights according to Bayes' formula \eqref{posterior} (blue) and using the proposed Morse potential \eqref{synder2}  (red). The state space
dimension is $N_x=10,30,$ and $N_x=100$ as in \cite{snyder2008obstacles}. The sample size is $N_e=10^3$ and the weight parameter $\alpha=\frac12.$} \label{fig1} 
\end{figure}

To illustrate the impact of the added potential to the weight computation in the particle filter, we consider the example suggested in Sec.~3 of \cite{snyder2008obstacles}. For convenience of the reader, we repeat the setup of the problem using the notation of \cite{snyder2008obstacles}. We consider a state $x \in \mathbb{R}^{N_x}$ for $N_x \in \{10, 30, 100 \}$. The observation model is given by $y=x + \epsilon$ where $\epsilon$ and $x$ are both normally distributed with zero mean and unit variance. The ensemble size is set to $N_e=10^3.$ For a classical particle filter, the weight of the posterior of the ensemble member $i$  is given by Equation~(3) of \cite{snyder2008obstacles} or our \eqref{posterior}, respectively. 
In this example, each $x_i^f$ is sampled from a Gaussian distribution  with unit variance. The likelihood is a multivariate Gaussian distribution with mean $y-x$ and unit variance as in the example in Sec.~3 of \cite{snyder2008obstacles}. 
Reported is the histogram of $\max\limits_{i=1,\dots,N_e} w_i$ in Figure \ref{fig1} (blue part of the histogram). As in \cite{snyder2008obstacles} we observe the clustering of weights at weight $w\approx 1$ that becomes particularly pronounced as the dimension $N_x$ increases, noting that $N_e$ remains fixed.
\par 
For the modified particle filter in \eqref{posteriori modified} and \eqref{modified weights}, the weights of the posterior are computed with the weighted potential $U'(x)=\left(1-e^{-|x|}\right)^2-1,$ i.e., for $i=1,\dots,N_e$ we consider 
\begin{align}\label{synder2}
{w}_i = \frac{ p(y|x_i^f) }{ \sum\limits_{j=1}^{N_e} p(y|x_j^f) }, \quad 
w_i(\alpha) = {w}_i - \alpha \frac{1}{N_e}\sum\limits_{j=1}^{N_e}  U'(w_j-w_i). 
\end{align}
The weight $\alpha>0$ is fixed in all subsequent computations to $\alpha=\frac12.$ The potential used is the Morse potential, i.e., 
\begin{align}\label{Morse}
    U(z) = - \frac12 \exp\left( - \frac12 |z| \right).
\end{align}
Again, for different dimensions $N_x,$ a histogram of the maximum weights is reported in Figure \ref{fig1} (orange). As expected, the additional forcing due to the potential $U$ leads to a more uniform weight distribution, and helps to offset the weight collapse at weight $w \approx 1$ observed in the classical particle filter as the dimension $N_x$ increases.

\clearpage

\subsection{Ensemble Kalman filters}\label{sec:ensemble kalman filter}
\label{II.D}
To conclude this section, we introduce another class of Monte Carlo based data assimilation algorithms known as ensemble Kalman filters \citep{evensen1994sequential,burgers1998analysis,evensen2022data}, which we will compare with in our numerical experiment with a nonlinear dynamical model. Ensemble Kalman filters (EnKFs) are an extension of the Kalman filter formalism introduced by \cite{kalman1960new}, where Kalman filters apply Bayes' theorem to assimilate observations assuming that (i) the prior distribution and likelihood are both Gaussian, and (ii) both the observation operator $H$ in \eqref{eq:observation def} and the model dynamics $f$ in \eqref{eq:model dynamics} are linear. The posterior distribution is therefore a Gaussian distribution whose mean and covariance can be computed explicitly \cite{kalman1960new}. Ensemble Kalman filters extend this formalism by allowing for nonlinear observation operators, nonlinear model dynamics, and by approximating covariances using Monte Carlo techniques. This is approach is done by updating an ensemble of states using the same Kalman filter formalism. For an ensemble of states $x_i^f \in \mathbb{R}^{N_x}$ for $i=1,2,\dots,N_e$ evolved (forecasted) by \eqref{eq:model dynamics}, the ensemble is updated to generate a new ensemble $x_i^a \in \mathbb{R}^{N_x}$ for $i=1,2,\dots,N_e$ given an observation $y$ defined by \eqref{eq:observation def}, 
\begin{subequations}
  \begin{align}
    x_i^a &= x_i^f + \mathbf{K}\left(y - (\mathbf{H}x_i^f + \eta_i)\right), \quad i=1,2,\dots,N_e,\label{eq:enkf update}\\
    \mathbf{K} &= \mathbf{P}^f\mathbf{H}\transpose\big(\mathbf{HP}^f\mathbf{H}\transpose+\mathbf{R}\big)^{-1}. \label{eq:enkf kalman gain}
  \end{align}
\end{subequations}
The matrix $\mathbf{K}$ is referred to as the Kalman gain. 
The $N_x\times N_y$ matrix $\mathbf{H}$ is a linearization of the observation operator $H$ in \eqref{eq:observation def}, the $\eta_i$ are independent samples from a Gaussian distribution with mean zero and covariance $\mathbf{R}$, and the $N_x\times N_x$ matrix $\mathbf{P}^f$ is the (forecast) error covariance matrix estimated from the ensemble $\{x_i^f\}$ using the empirical sample covariance estimator,
\begin{equation}\label{eq:sample covariance definition}
    \mathbf{P}^f = \frac{1}{N_e-1}\sum_{i=1}^{N_e}(x_i^f-\overline{x}^f)(x^f_i-\overline{x}^f)\transpose,\quad
    \bar{x}^f = \frac{1}{N_e}\sum_{i=1}^{N_e} x_i^f.
\end{equation}
Equation~\ref{eq:enkf update} defines the ``stochastic'' EnKF \cite{burgers1998analysis}, where the term ``stochastic" arises from generating an ensemble of observations through the addition of $\eta_i$, which ensures the correct error statistics \citep{burgers1998analysis}.
Variants of the EnKF include the ensemble adjustment Kalman filter \citep{anderson2001ensemble} and ensemble transform filters \citep{tippett2003ensemble}, for example. 

Unlike particle filters, the EnKF assumes that the prior distribution and likelihood are both Gaussian in order to apply the Kalman filter formalism to update the forecasted particles in \eqref{eq:enkf update}. 
Ensemble Kalman filters, therefore rely on the accuracy of the mean state and covariance estimates from the empirical estimators in \eqref{eq:sample covariance definition}. The empirical sample covariance in \eqref{eq:sample covariance definition} is unbiased and consistent, however produces large errors when $N_e \ll N_x$ \citep{wainwright2019high}. In these contexts, additional covariance estimation techniques are employed to mitigate these errors due to limited ensemble size, for example by reducing spurious correlations through tapering \citep[][Ch.~10 and references therein]{houtekamer2001sequential,evensen2022data}. 
Since we are interested in comparing the EnKF with the particle filter, we will consider the case where $N_e \gg N_x$ so that we can compute \eqref{eq:sample covariance definition} directly with sufficient accuracy.


\section{Numerical experiments for linear dynamics}\label{III}
To assess the performance of the modified particle filter in a cycling data assimilation context, we first test the method for a linear dynamical model in which the true state is known. 
In this example, we consider a simple time-discrete linear model with parameter $\lambda>0$ and time step $\Delta t>0.$ The evolution of the true state is given by 
\begin{align}\label{linear true}
    x_{true}^{n}=x^0 \; \exp(- \lambda \; \Delta t \; n ). 
\end{align}
for an initial data $x^0_{true} \in \mathbb{R}.$ The time horizon is $n \Delta t = 2.$ The observation model is 
\begin{align}\label{noise linear}
    y = x_{true} + \sigma \epsilon 
\end{align}
where $\epsilon$ is the normally distributed noise with unit variance and the positive parameter $\sigma>0$. Observations are recorded at each time $n=0,1 \dots.$ The dynamics for the state estimate \eqref{eq:model dynamics} is defined by a time--discrete forward model with states $x^f_n$ at time $n \Delta t.$ As forward model we use an explicit Euler discretization, i.e., 
\begin{align} \label{linear forecast}
    x^{f,n+1} = x^{f,n} - \lambda \Delta t \;  x^{f,n}.
\end{align}
 The same time step as above is used and the initialization $x^{f,0}$ is distributed according to a prior distribution $p(x)$ specified below. 

We compare the performance of the classical particle filter and modified particle filter for estimating the current state of the system at time $n \Delta t.$ The initial data $x^{f,0}$ is distributed according to $p$, which is taken to be a normal distribution with mean $\bar{x}^{f,0}$ and variance $\sigma^f.$ An iterative procedure is applied to identify the true state $x_{true}^n.$ At each time step, we apply Bayes' theorem and compute the weights $w_i^n$ using equation \eqref{weights} or equation \eqref{modified weights}, respectively. The posteriori distribution $p(x^{f,n}|y)$ is given by equation \eqref{posterior} or equation \eqref{posteriori modified}, respectively. The likelihood is a multivariate Gaussian distribution with mean $y-x^{f,n}$ and unit variance. For the next time step $n+1,$ this posteriori is re-sampled to obtain a new prior of the type given by equation \eqref{dirac}. At each point in time we record the mean and variance of the posterior $p(x^{f,n}|y)$ and we compare this numerically to the true state $x_{true}^n.$ 

\begin{figure}[htb]
\center
\includegraphics[width=.45\textwidth]{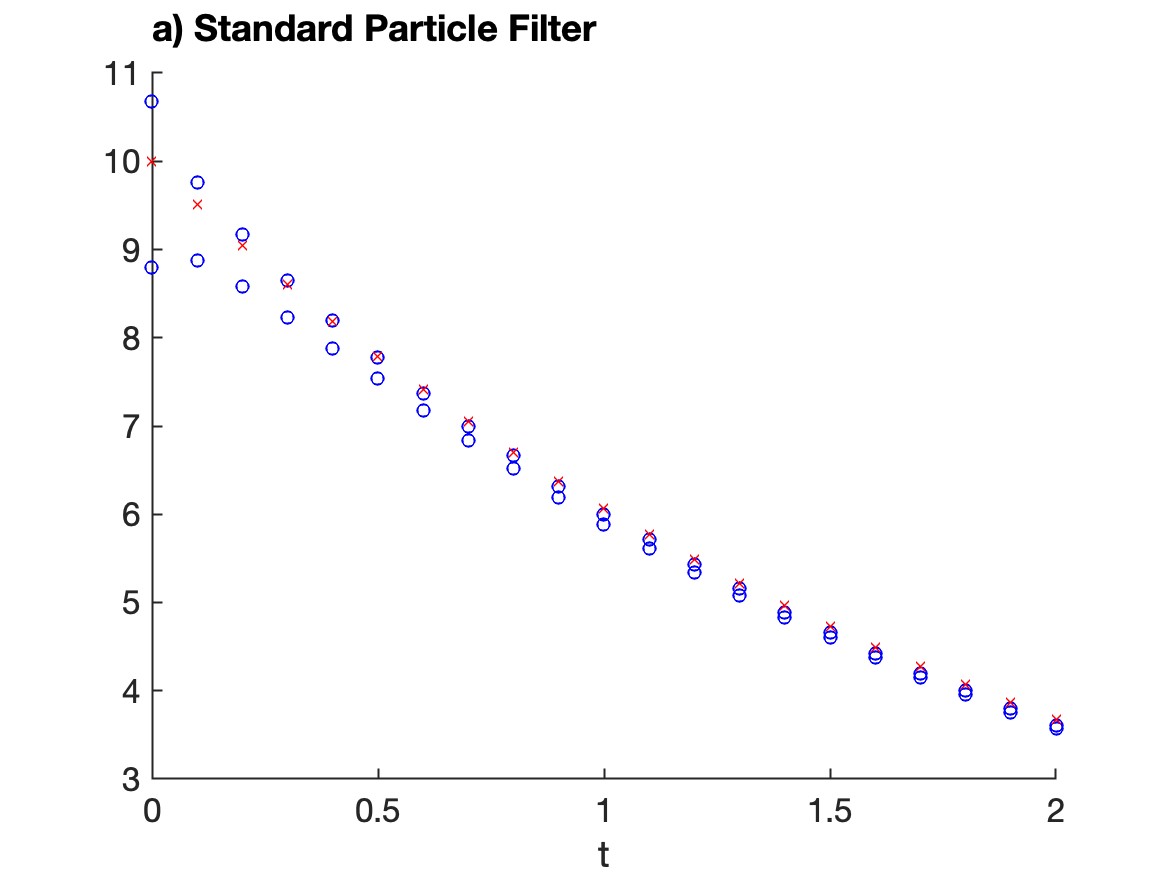}
\includegraphics[width=.45\textwidth]{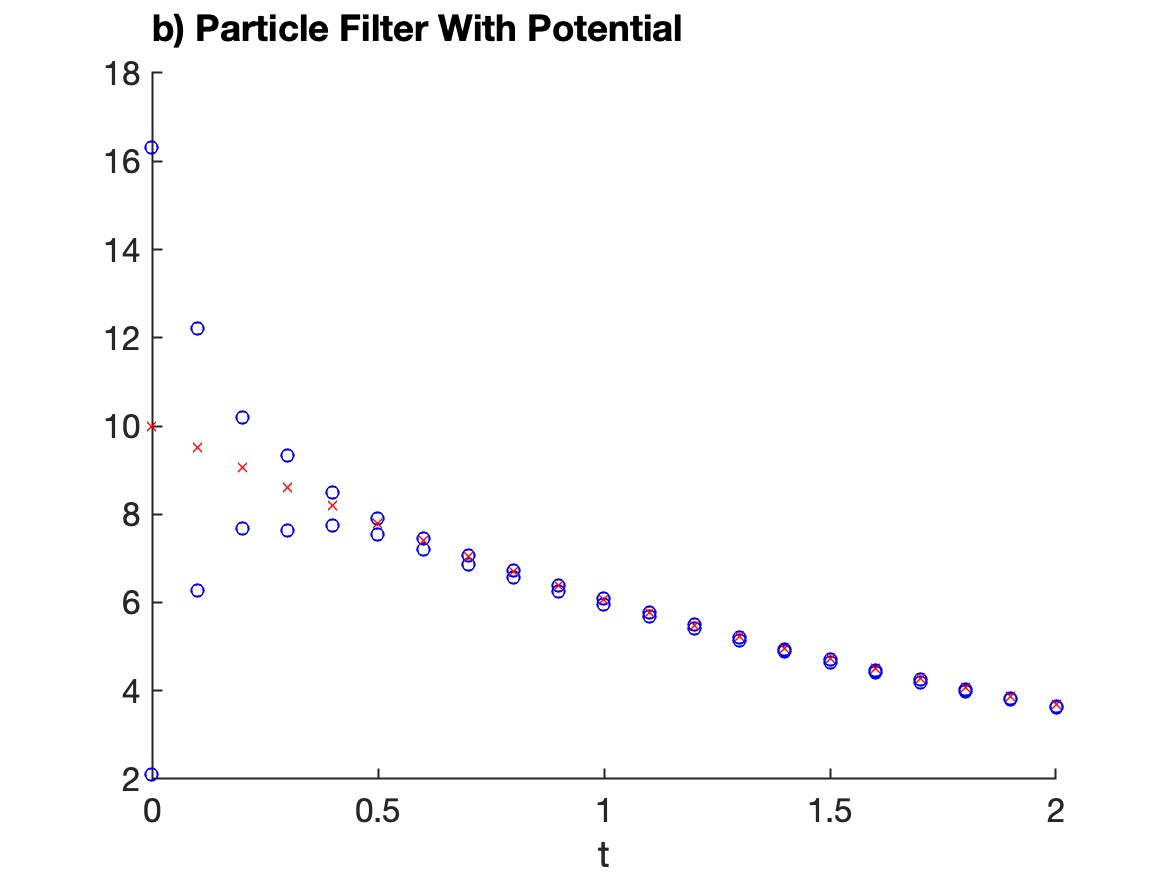}
\caption{Particle filter for the linear model \eqref{linear true}, with forecast model \eqref{linear forecast}. Noise is added to the observation according to \eqref{noise linear}. In red the true solution is depicted. In blue, we present the mean of the ensemble where we add and subtract the variance of the ensemble. In part a) of the figure the classical particle filter defined through the weights \eqref{weights} is shown. In  part b) of the figure the modified particle filter defined through \eqref{modified weights}.  
} \label{figlin}
\end{figure}

We expect that the variance of the particle filter decays for increasing number $n$ as well as convergence of the mean of the posterior towards the true state. Since in the modified version of the particle filter \eqref{posteriori modified} and \eqref{modified weights} we modify the weights, we expect a slower decay of the variance for the modified algorithm. In both cases, we observe the expected decay towards the true mean in Figure \ref{figlin}. The modification applied to the computation of the weights \eqref{modified weights}  is expected to introduce an additional variance to the sample. The example shows that this additional variance does not deter the overall properties of the particle filter, namely, the convergence to the true state. In the right part of Figure \ref{figlin} it is shown that even so the variance is larger on the first time steps, it decays as in the classical particle filter (left part). 
\par 
The following parameters are used in numerical simulations: $\Delta t = \frac{1}{10}$, $\lambda=\frac12$, $\sigma=\frac1{10},$ $x^0_{true}=10$,  $\bar{x}^{f,0}=8$  and $\sigma^f=4.$ We use $N_e=10^3$ particles and a weight $\alpha=10^{-3}.$ The potential is the same as in the first example given by equation \eqref{Morse}. The number of time steps for the simulation is $T/\Delta t.$

\section{Numerical experiments with Lorenz '63}\label{IV}
We extend the linear example of the previous section to a nonlinear dynamical system. In this example, we compare the classical particle filter with the particle filter with modified weights and the ensemble Kalman filter (EnKF) in a series of cycling data assimilation experiments with the Lorenz '63 dynamical model \citep{lorenz1963deterministic}. The Lorenz '63 dynamical model is system of three nonlinear, coupled ordinary differential equations for scalar variables $x(t),y(t)$ and $z(t)$,
\begin{align}
    x' &= \sigma(y-x), \nonumber \\ 
    y' &= \rho x - xz-y, \label{eq:lorenz63} \\
    z' &= xy-bz. \nonumber
\end{align}
We fix the parameters $\sigma=10,\rho=28,b=8/3$, and define the transpose of the state vector $\mathbf{x}^\text{T} = [x \ y \ z ]^\text{T}$.

The set up for these experiments is similar to that described in Sec.~4b of \cite{evensen2000ensemble}. We define our reference solution $\mathbf{x}_\text{ref}$ as the solution to \eqref{eq:lorenz63} approximated by a fourth-order explicit Runge-Kutta scheme with time step $\Delta t = 0.01$ and initial condition
\begin{equation}
    \mathbf{x}_{\text{ref}}^0= \left[\begin{array}{r} 1.508870 \\ -1.531271 \\ 25.46091\end{array}\right].
\end{equation}
We generate the initial ensemble of states $\mathbf{x}^0_{i}$, $i=1,2,\dots,N_e$ from the reference initial condition $\mathbf{x}_{\text{ref}}^0$ by adding independent and identically distributed (IID) noise that is Gaussian distributed with mean zero and variance two.
The observations $y$ are generated from reference solution using \eqref{eq:observation def}
with an observation operator $H = \mathbf{H} = [1 \ 0 \ 0]$ and $\epsilon$ are IID Gaussian with mean zero and observation error variance $r=1$. Observations are assimilated every $10$ integration time steps (i.e., every $10\Delta t$). Assimilation experiments are run for a total of 100 cycles, starting each cycle with a filter update, followed by a forecasting step. For the forecasting step, we evolve $i=1,2,\dots,N_e$ particles according to
\begin{equation}
    \mathbf{x}^{k+1}_{i} = f(\mathbf{x}^k_{i}) + \bs{\omega}^k,
\end{equation}
where $f(\cdot)$ is the same explicit Runge-Kutta scheme used to generate the reference solution with index $k$ denoting one integration time step $\Delta t$. The vector $\bs{\omega}$ is a stochastic forcing term which we define to be
\begin{equation}
    \bs{\omega} := \sigma \sqrt{\Delta t}\left[\begin{array}{r}\sqrt{2} \\ \sqrt{12.13} \\ \sqrt{12.31} \end{array}\right],
\end{equation}
where $\sigma$ are independent draws from a standard normal distribution. This corresponds to adding uncorrelated model errors with a diagonal model error matrix $\mathbf{Q} = \text{diag}(2,12.13,12.31)$.

In these experiments, we compare the state estimates after a filter update, which we refer to as the analysis state, from the classical particle filter, modified particle filter, and EnKF. We consider two ensemble sizes of $N_e=50$ and $N_e=100$.
For both particle filters we apply (uniform) sequential importance resampling to generate the analysis particles \citep[][p.~1993]{van2010nonlinear}. 
Therefore, the only difference between the two particle filters is in the calculation of the weights. The stepsize $\alpha>0$ in the modified particle filter is chosen via a line search, where for a fixed number of particles $N_e$, we choose the optimal $\alpha$ that minimizes the average root mean square (RMS) error in the analysis state over a series of 500 data assimilation experiments. 

\begin{figure}[htb]
    \centering
    \includegraphics[width=0.8\linewidth]{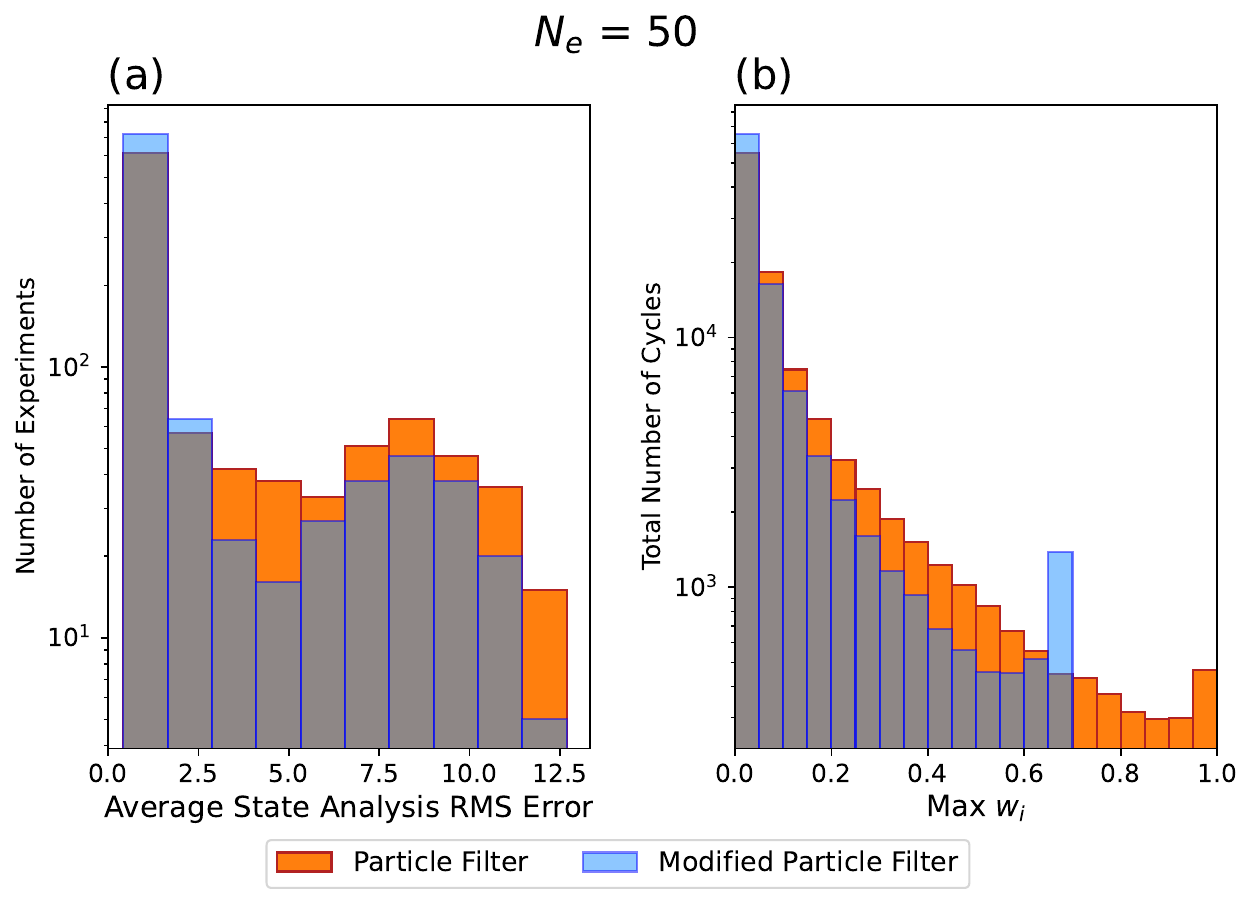}
    \caption{Panel (a): Average state analysis root mean square (RMS) errors over 1000 cycling experiments with the Lorenz '63 model and $N_e=50$. Panel (b): Histogram of the maximum weight $w_i$ of each cycle for the 1000 cycling experiments. Orange corresponds to the classical particle filter, and blue the modified particle filter. Vertical scales in both panels are logarithmic scales.}
    \label{fig:fig3}
\end{figure}

\begin{figure}[htp]
    \centering
    \includegraphics[width=0.8\linewidth]{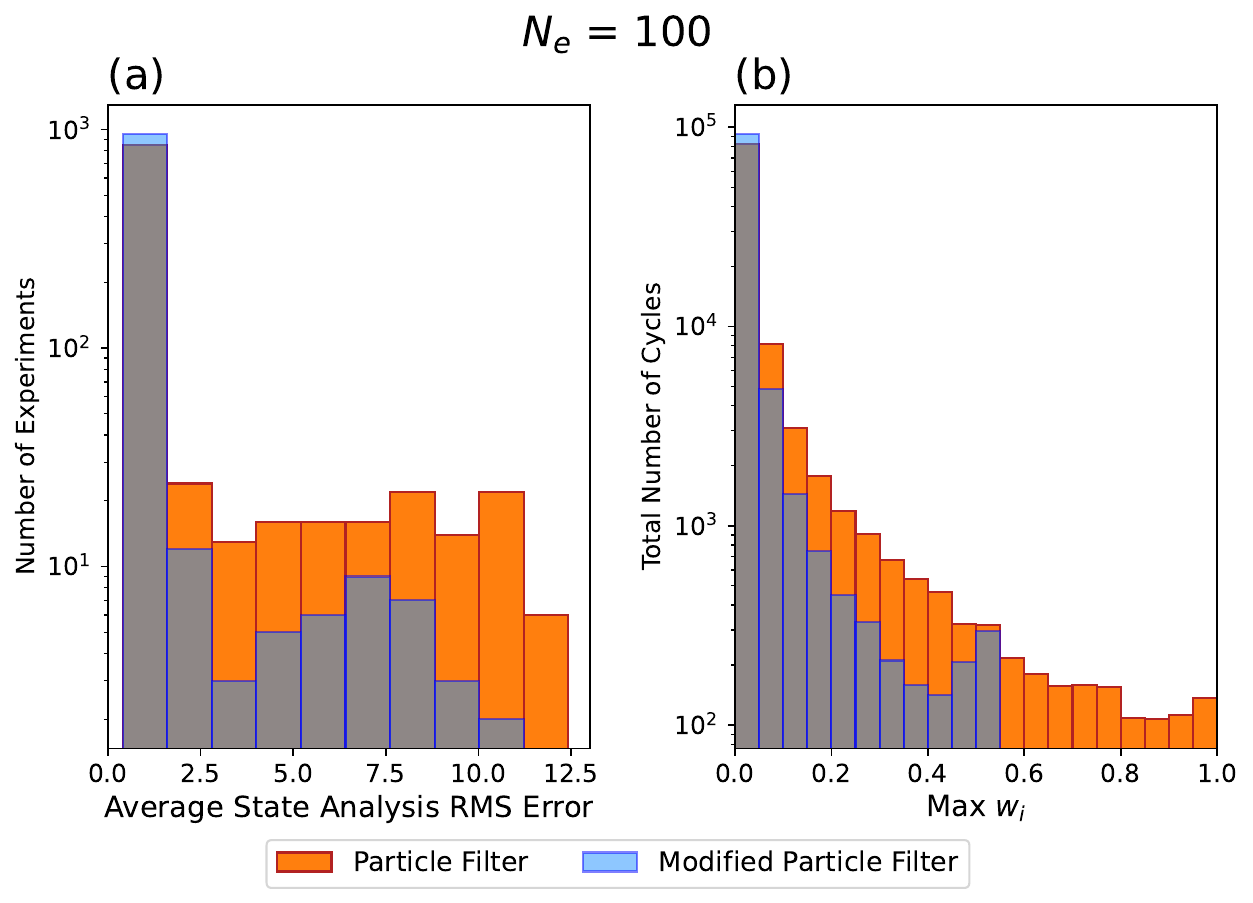}
    \caption{Same as Fig.~\ref{fig:fig3} for $N_e=100$.}
    \label{fig:fig4}
\end{figure}

Figures~\ref{fig:fig3} and \ref{fig:fig4} summarize the results for 1000 cycling experiments with the Lorenz '63 model for $N_e=50$ and $N_e=100$, respectfully. Panel (a) of Figs.~\ref{fig:fig3} and \ref{fig:fig4} are histograms of the average state analysis root mean square (RMS) error for all these experiments, where in each experiment the first 30 cycles are removed from the averaging for ``spin-up." We see that the modified particle filter generally reduces the state analysis RMS error, in particular reducing the frequency of experiments with large RMS error. This is reflected in the mean, median, and standard deviation of these averaged state analysis errors in Table~\ref{table:lorenz state errors}. We use the average analysis RMS error from the EnKF as a reference for comparison with the two particle filters. Even though the statistics over the 1000 experiments indicate larger errors in both particle filters relative to the EnKF, individual experiments have state analysis RMS that are comparable to that of the EnKF.
The modified particle filter overall improves the mean state analysis error relative to the classical particle filter, and reduces the standard deviation in these errors, most notably in the $N_e=100$ case. While the median state RMS error for the modified particle filter is larger than the classical particle filter in the $N_e=100$ case, the magnitude of the large errors is reduced, as reflected in the lower standard deviation and histogram in Fig.~\ref{fig:fig4}. 

\begin{figure}[htp]
    \centering
    \includegraphics[width=0.8\linewidth]{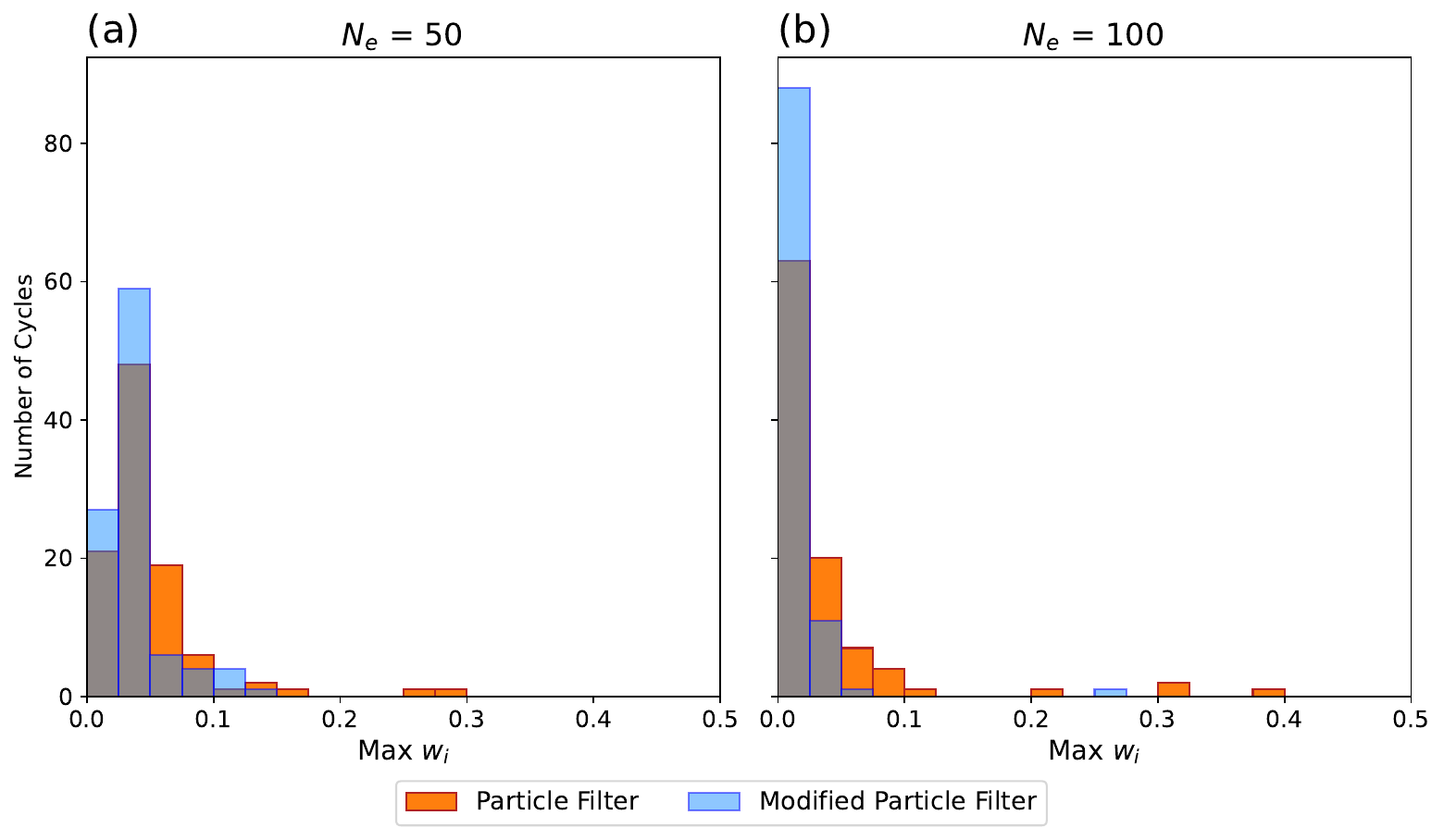}
    \caption{Maximum weights $w_i$ for the classical particle filter (orange) and modified particle filter (blue) over a single cycling experiment of 100 cycles. Panel (a) corresponds to $N_e=50$ and panel (b) $N_e=100$. The average state analysis RMS errors (removing the first 30 cycles) for the classical particle filter is 0.506 for $N_e=50$ and 0.608 for $N_e=100$; for the modified particle filter 0.424 for $N_e=50$ and 0.568 for $N_e=100$.}
    \label{fig:fig5}
\end{figure}

The improvements in the state estimates from the modified particle filter can be related to the improved distributions of weights. Panel (b) of Figs.~\ref{fig:fig3} and \ref{fig:fig4} plots histograms of the maximum weights $w_i$ for each of the 100 cycles and over all 1000 experiments. As observed in the motivating example, the modified particle filter shifts the distribution of maximum weights towards smaller values, reducing the instances of weight collapse to values of one. 
This is also reflected in the weight distributions for a single experiment in Fig.~\ref{fig:fig5}, in which the modified particle filter has slightly smaller state analysis RMS errors than the classical particle filter.

\begin{table}
\caption{State analysis RMS Error over 1000 experiments for the classical particle filter (PF), modified particle filter (PF), and ensemble Kalman filter (EnKF). \label{table:lorenz state errors}}
\begin{ruledtabular}
\begin{tabular}{@{} lcccccc @{}} 
& \multicolumn{3}{c}{$N_e=50$} & \multicolumn{3}{c}{$N_e=100$}\\
         & PF & mPF & EnKF & PF & mPF & EnKF\\ \hline
      Mean & 3.077& 2.375& 0.642 & 1.584&1.030 &0.635 \\ \hline
      Median & 1.066&0.939 & 0.620 & 0.699&0.782 &0.626 \\ \hline
      Standard Deviation &3.393 & 2.931& 0.104 & 2.431&1.170 &0.102\\ \hline
   \end{tabular}
\end{ruledtabular}
\end{table}

In addition to the distributions of the maximum weights, we also compute two additional quantities for comparison. The first is the variance in the weights, $\tau^2$,
\begin{equation}
    \tau^2 = \text{var}\log(w),
\end{equation}
derived in Sec. 4a of \cite{snyder2008obstacles}. Larger values of $\tau$ correspond to higher variance in the weights and can be indicative of collapse. Table~\ref{table:tau} summarizes the mean, median, and standard deviation of $\tau$ for the classical particle filter, and modified particle filter over the 1000 experiments with $N_e=50$ and $N_e=100$. For both ensemble sizes, the modified particle filter reduces the variances in the computed weights and significantly lowers the standard deviation. As a consequence, the modified particle filter decreases the frequency of experiments with large state analysis errors.
We also see that the modified particle filter reduces the effective ensemble size,
$N_\text{eff}$,
\begin{equation}
    N_\text{eff} = \bigg(\sum_{i=1}^{N_e}w_i^2\bigg)^{-1},
\end{equation}
\citep[][their (9.5)]{evensen2022data}, given in Table~\ref{table:effective ensemble size}. Small effective ensemble sizes $N_\text{eff}$ indicate collapse and the need for resampling. The modified particle filter increases the effective ensemble size relative to the classical particle filter, particularly in the $N_e=100$ case.

\begin{table}
\caption{Variance in the computed weights ($\tau$) over 1000 experiments for the classical particle filter (PF) and modified particle filter (PF).\label{table:tau}}
\begin{ruledtabular}
    \begin{tabular}{@{} lcccc @{}} 
    & \multicolumn{2}{c}{$N_e=50$} & \multicolumn{2}{c}{$N_e=100$}\\
      & PF & mPF & Pf &mPF \\ \hline
      Mean & 5.558&0.316 & 5.141&0.164\\ \hline
      Median & 0.837&0.252 & 0.596&0.123 \\ \hline
      Standard Deviation & 19.39& 0.246&22.26 &0.130\\ \hline
   \end{tabular}
\end{ruledtabular}
\end{table}

\begin{table}
\caption{Effective ensemble size ($N_\text{eff}$) for the 1000 experiments for the classical particle filter (PF) and modified particle filter (PF).\label{table:effective ensemble size}}
\begin{ruledtabular}
    \begin{tabular}{@{} lcccc @{}} 
    & \multicolumn{2}{c}{$N_e=50$} & \multicolumn{2}{c}{$N_e=100$}\\
    & PF & mPF & PF & mPF\\ \hline
      Mean & 31& 36 & 69& 82\\ \hline
      Median & 36& 41& 79& 90\\ \hline
      Standard Deviation & 16& 14&28 &20 \\ \hline
    \end{tabular}
\end{ruledtabular}
\end{table}

\section{Summary and discussion}\label{sec:summary}\label{V}
Particle filters are a class of Monte Carlo data assimilation techniques that are known to suffer from weight collapse, or weight degeneracy, where the weight of one of the particles concentrates at one while all others become very close to zero. We introduce a small modification to the particle filter, inspired by recent developments in energy-based diversity measures, to mitigate this collapse. The proposed modification introduces a potential on the weight distribution of the filter and adjusts the weight values based on a minimization of the corresponding potential. Through a series of numerical experiments with linear and nonlinear dynamics, we compare the performance of this new, modified particle filter, with the classical particle filter, and with the ensemble Kalman filter in the nonlinear case. We find that this modification helps to reduce weight collapse by improving weight distributions while still yielding the correct Bayes's posterior distribution. In the numerical experiments with the nonlinear Lorenz '63 model, we find both improved state estimates and reduced frequency of data assimilation experiments with significant errors when including the potential function in the weight computations.

The findings in our numerical experiments suggest that modifying the weight computation with a potential function can result in improved particle filter performance for a moderately small number of particles. In future work we plan to investigate the numerical impact of different potential functions, like e.g. the Riesz potential, as well as analytical considerations on the large particle limit of the proposed method.

 



\medskip 

{\small \noindent {\bf Acknowledgments}
MH is funded by the Deutsche Forschungsgemeinschaft (DFG, German Research Foundation) - SPP 2183:  Eigenschaftsgeregelte Umformprozesse with the Project(s)  HE5386/19-3 Entwicklung eines flexiblen isothermen Reckschmiedeprozesses für die eigenschaftsgeregelte Herstellung von Turbinenschaufeln aus Hochtemperaturwerkstoffen (424334423) and through 
HE5386/33-1 Control of Interacting Particle Systems, and Their Mean-Field, and Fluid-Dynamic Limits (560288187), HE5386/34-1 Partikelmethoden für unendlich dimensionale Optimierung ( 561130572) and  through 442047500/SFB1481 within the projects B04 (Sparsity fördernde Muster in kinetischen Hierarchien), B05 (Sparsifizierung zeitabhängiger Netzwerkflußprobleme mittels diskreter Optimierung) and B06 (Kinetische Theorie trifft algebraische Systemtheorie) is acknowledged. SG is supported in part by the Data Driven Discovery RTG at the University of Arizona under NSF Grant No. DMS-1937229.
 }

 {\small \noindent {\bf Author Contributions}
S.G. and M.H. contributed equally to this work.
 }

\bibliography{references}

\end{document}